\documentclass[10pt,a4paper]{article}

\usepackage{authblk}
\usepackage[T1]{fontenc}
\usepackage{graphicx}
\usepackage{hyperref}

\hypersetup{
  colorlinks   = true, %Colours links instead of boxes
  urlcolor     = blue, %Colour for external hyperlinks
  linkcolor    = blue, %Colour of internal links
  citecolor   = blue %Colour of citations
}
\usepackage{listings}
\usepackage{amsmath}
\usepackage{amssymb}
\usepackage{siunitx}

\makeatletter
\AtBeginDocument{%
  \@ifpackageloaded{hyperref}
  {\def\@doi#1{\href{https://doi.org/#1}
      {\ttfamily https://doi.org/#1}\egroup}}
  {\def\@doi#1{\ttfamily https://doi.org/#1\egroup}}
  \def\doi{\bgroup\catcode`\_=12\relax\@doi}}
\makeatother
\newcommand{\SMPT}{\textsc{SMPT}}

\providecommand{\keywords}[1]
{
  \small	
  \textbf{\textit{Keywords---}} #1
}

\title{{SMPT}: A Testbed for Reachability Methods\\ in Generalized Petri Nets}
\author[1]{Nicolas Amat}
\author[1]{Silvano Dal Zilio} 
\affil[1]{LAAS-CNRS, Universit\'{e} de Toulouse, CNRS, Toulouse, France}
\date{}
\begin{document}

\maketitle              
\sloppy

\begin{abstract}
SMPT (for Satisfiability Modulo Petri Net) is a model checker for reachability
problems in Petri nets. It started as a portfolio of methods to experiment
with symbolic model checking, and was designed to be easily extended. Some
distinctive features are its ability to benefit from structural reductions and
to generate verdict certificates. Our tool is quite mature and performed well
compared to other state-of-the-art tools in the Model Checking Contest.

\end{abstract}
%
%
%%%%%%%%%%%%%%%%%%%%%%%%%%%%%%%%%%%%%%%%%%%%%%%%%%%%%%%%%%%%%%%%%%%%%%%%%%%%%%%
%%%%%%%%%%%%%%%%%%%%%%%%%%%%%%%%%%%%%%%%%%%%%%%%%%%%%%%%%%%%%%%%%%%%%%%%%%%%%%%

\section{Introduction}

\SMPT{} is an open source model checker designed to answer reachability queries
on generalized Petri nets, meaning that we do not impose any restrictions on the
marking of places or the weight on the arcs. We can in particular handle
unbounded nets. We also support a generalized notion of {reachability
properties}, in the sense that we can check if it is possible to reach a marking
that satisfies a combination of linear constraints between places. This is more
expressive than the reachability of a single marking and corresponds to the
class of formulas used in the reachability category of the Model Checking
Contest (MCC), a yearly competition of formal verification tools for concurrent
systems~\cite{abcdg2019,khhjp2021}.

The tool name is an acronym that stands for \emph{Satisfiability
  Modulo Petri Net}. This choice underlines the fact that, for most of the
new features we implemented, \SMPT{} acts as a front-end to a SMT
solver; but also that it adds specific knowledge from Petri net
theory, such as invariants, use of structural properties, etc. %\\

The design of \SMPT{} reflects the two main phases during its
development process. The tool was initially developed as a testbed for
symbolic model checking algorithms that can take advantage of
structural reductions (see e.g.~\cite{pn2021,Amat-FI-22}). This explains why it
includes many ``reference'' implementations of fundamental
reachability algorithms, tailored for Petri nets, such as Bounded
Model Checking
(BMC)~\cite{clarke_bounded_2001,heljanko2001bounded,armando_bounded_2006}
or $k$-induction~\cite{sheeran_checking_2000}. It also includes new
verification methods, such as adaptations of Property Directed
Reachability
(PDR)~\cite{jhala_sat-based_2011,hutchison_understanding_2012} for
Petri nets~\cite{tacas2022}. One of our goal is to efficiently compare different
algorithms, on a level playing field, with the the ability to switch
on or off optimizations. This motivates our choice to build a tool that
is highly customizable and easily extensible.

In a second phase, since 2021, we worked to make \SMPT{} more mature,
with the goal to improve its interoperability, and with the addition
of new verification methods that handle problems where symbolic
methods are not the best suited. We discuss the portfolio approach
implemented in \SMPT{} in Sect.~\ref{sec:design}. This second set of
objectives is carried by our participation in the last two editions of
the MCC~\cite{mcc:2021,mcc:2022}, where we obtained a $100\%$
confidence level (meaning \SMPT{} never returned an erroneous
verdict). With this last evolution, we believe that \SMPT{} left its
status of prototype to become a tool that can be useful to other
researchers. This is what motivates the present paper. %\\

There are other tools that perform similar tasks. We provide a brief
comparison of \SMPT{} with two of them in Sect.~\ref{sec:mcc},
\textsc{ITS-Tools}~\cite{its_tools,thierry-mieg_structural_2020} and
\textsc{Tapaal}~\cite{tapaal}. All tools have in common their
participation in the MCC and the use of symbolic techniques. They also share
common input formats for nets and formulas. We can offer two reasons
for users to use \SMPT{} instead of---or more logically in addition
to---these tools. First, \SMPT{} takes advantage of a new approach,
called \emph{polyhedral reduction}~\cite{pn2021,Amat-FI-22}, to accelerate the
verification of reachability properties. This approach can be
extremely effective in some cases where other methods do not scale. We
 describe this notion in Sect.~\ref{sec:theory}. Another
interesting feature of \SMPT{} is the ability to return a
\emph{verdict certificate}. When a property is invariant, we can
return a ``proof'' that can be checked independently by a SMT
solver.

%%%%%%%%%%%%%%%%%%%%%%%%%%%%%%%%%%%%%%%%%%%%%%%%%%%%%%%%%%%%%%%%%%%%%%%%%%%%%%%
%%%%%%%%%%%%%%%%%%%%%%%%%%%%%%%%%%%%%%%%%%%%%%%%%%%%%%%%%%%%%%%%%%%%%%%%%%%%%%%

\section{Technical Background}
\label{sec:theory}

We briefly review some theoretical notions related to our work. We
assume basic knowledge of Petri net theory~\cite{reisig2012petri}. In
the following, we use $P$ for the set of places of a net $N$. A
marking, $m$, is a mapping associating a non-negative integer, $m(p)$,
to every place $p$ in $P$. \SMPT{} supports the verification of
\emph{safety properties} over the reachable markings of a marked Petri
net $(N, m_0)$. Properties, $F$, are defined as a Boolean combination
of literals of the form $\alpha \sim \beta$, where $\sim$ is a
comparison operator (one of $=$, $\leqslant$ or $\geqslant$) and
$\alpha$, $\beta$ are linear expressions involving constants or places
in $P$. For instance, $(p + q \geqslant r) \vee (p \leqslant 5)$ is an
example property.

We say that property $F$ is valid at marking $m$, denoted
$m \models F$, if the ground formula obtained by substituting places,
$p$, by $m(p)$ is true. As can be expected, we say that $F$ is
\emph{reachable} in $(N, m_0)$ if there is $m$ reachable such that
$m \models F$. See~\cite{pn2021,Amat-FI-22,tacas2022} for more
details.  We support two categories of queries: $\text{EF} \, F$,
which is true only if $F$ is reachable; and $\text{AG} \, F$, which is
true when $F$ is an invariant, with the classic relationship that
$\text{AG}\, F \equiv \neg \, (\text{EF}\, \neg F)$. A \emph{witness}
for property $\text{EF} \, F$ is a reachable marking such that
$m \models F$; it is a \emph{counterexample} for
$\text{AG} \, \neg F$. Examples of properties we can express in this
way include: checking if some transition $t$ is enabled
(quasi-liveness); checking if there is a deadlock; checking whether
some linear invariant between places is always true; etc.

\SMPT{} implements several methods that combine SMT-based techniques
with a new notion, called polyhedral reduction. The idea consists in
computing structural
reductions~\cite{berthelot_transformations_1987,berthomieu_counting_2019}
of the form $(N_1, m_1) \vartriangleright_E (N_2, m_2)$, where
$(N_1, m_1)$ is the (initial) Petri net we want to analyse;
$(N_2, m_2)$ is a reduced version; and $E$ is a system of linear
equations relating places in $N_1$ and $N_2$. The goal is to preserve
enough information in $E$ so that we can reconstruct the reachable
markings of $(N_1, m_1)$ by knowing only those of $(N_2, m_2)$. Given
a starting net, we can automatically compute a polyhedral reduction
using the tool \textsc{Reduce}, which is part of
\textsc{Tina}~\cite{berthomieu2004tool}. (But obviously there are many
irreducible nets.)

Polyhedral reductions are useful in practice. Given a property $F_1$ on the
initial net $N_1$, we can build a property $F_2$ on
$N_2$~\cite{pn2021,Amat-FI-22} such that checking $F_1$ on $N_1$ (whether it is
reachable or an invariant) is equivalent to checking $F_2$ on
$N_2$. We have observed very good speed-ups with this approach, even
when we only have a moderate amount of reductions. This notion is also
``compatible'' with symbolic methods. In \SMPT{}, we recast all
constraints and relations into formulas of Quantifier Free Linear
Integer Arithmetic (the QF-LIA theory in the SMT-LIB
standard~\cite{smt-lib}) and pass them to SMT solvers.

Another important notion is that of \emph{inductive invariant}. We say
that $R$ is an inductive invariant of property $F$ if it is: (i) valid
initially ($m_0 \models R$); (ii) inductive (if $m \to m'$ and
$m \models R$ then $m'\models R$, for all markings $m$, even those
that are not reachable); and (iii) $R \supseteq F$. Given a pair
$(F, R)$ we can check these three properties automatically using a SMT
solver (and with only one formula in each case). In some conditions,
when property $F$ is an invariant, \SMPT{} can automatically compute
an inductive invariant from $F$. This provides %, in fact,
an independent certificate that invariant $F$ holds.

%%%%%%%%%%%%%%%%%%%%%%%%%%%%%%%%%%%%%%%%%%%%%%%%%%%%%%%%%%%%%%%%%%%%%%%%%%%%%%%
%%%%%%%%%%%%%%%%%%%%%%%%%%%%%%%%%%%%%%%%%%%%%%%%%%%%%%%%%%%%%%%%%%%%%%%%%%%%%%%

\section{Design and Implementation}
\label{sec:design}

\SMPT{} is open-source, under the GNU GPL v3.0 licence, and is freely
available on GitHub (\url{https://github.com/nicolasAmat/SMPT}). The
repository also provides examples of nets, formulas, and scripts to
experiment with the tool.
% Our examples are also used for benchmarking
% and for continuous testing.
\SMPT{} is a Python project of about
$4\,000$ lines of code, and is fully typed using the static type
checker \href{http://www.mypy-lang.org}{\texttt{mypy}}. The code is
heavily documented ($4\,500$ lines) and we provide many tracing and debugging options
that can help understand its inner workings. The project is packaged
in libraries, and provides abstract classes to help with future
extensions.
We describe each library and explain how they can be extended.
\smallbreak

\noindent\textbf{The ptio library} defines the main data-structures of the model
checker, for Petri nets (\texttt{pt.py}), reachability formulas
(\texttt{formula.py}), and reduction equations (\texttt{system.py}).
It also provides the corresponding parsers, for different formats.
\smallbreak

\noindent\textbf{The interface library} includes interfaces to external
tools and solvers. For example, we provide an integrated interface to
\textsc{z3}~\cite{z3} built around the SMT-LIB
format~\cite{smt-lib}. We can also interface with
\textsc{MiniZinc}~\cite{nethercote2007minizinc}, a solver based on
constraint programming techniques, and with a random state space
explorer, \textsc{walk}, distributed with the \textsc{Tina}
toolbox. New tools can be added by implementing the abstract class
\texttt{Solver} (\texttt{solver.py}).
\smallbreak

\noindent\textbf{The exec library} provides a concurrent ``jobs
scheduler'' that helps run multiple verification tasks in parallel and
manage their interactions. \smallbreak

\noindent\textbf{The checker library} is the core of our tool. It includes a
portfolio of methods intended to be executed in parallel. All methods
implement an abstract class (\texttt{abstractchecker.py}) which
describes the abstract method \texttt{prove}. We currently support the
following eight methods: \smallbreak

\noindent (1)\ \textbf{Induction}: a basic method that checks if a
property is an inductive invariant (see Sect.~\ref{sec:theory}). This
property is ``easy'' to check, even though interesting properties are
seldom inductive. It is also useful to check verdict certificates.\smallbreak

\noindent (2)\ \textbf{BMC}: Bounded Model Checking
\cite{biere_symbolic_1999} is an iterative method to explore the state
space of systems by unrolling their transitions. This method is only
useful for finding counterexamples.\smallbreak

\noindent (3)\ \textbf{$k$-induction}: \cite{sheeran_checking_2000} is
an extension of BMC that can also prove invariants. \smallbreak

\noindent (4)\ \textbf{PDR}: Property Directed Reachability
\cite{jhala_sat-based_2011,hutchison_understanding_2012}, also known
as IC3, is a method to strengthen a property that is not inductive,
into an inductive one. This method can return a verdict
certificate. We provide three different methods of increasing
complexity \cite{tacas2022} (one for coverability and two for general
reachability). \smallbreak
    
\noindent (5)\ \textbf{State Equation}: is a method for checking that
a property is true for all ``potentially reachable markings''
(solution of the state equation). This is a semi-decision method,
found in many portfolio tools, that can easily check for
invariants. We implement a refined
version~\cite{its_tools,thierry-mieg_structural_2020} that can
over-approximate the result with the help of trap constraints
\cite{trap_constraints} and other structural information, such as NUPN
specifications~\cite{garavel_nested-unit_2019}. \smallbreak

\noindent (6)\ \textbf{Random Walk}: relies on simulation tools to
quickly find counterexamples. It is also found in many tools that
participate in the MCC~\cite{khhjp2021}. We currently use
\textsc{walk}, distributed with the \textsc{Tina} toolbox, but we are
developing a new tool to take
advantage of polyhedral reductions. \smallbreak
    
\noindent (7)\ \textbf{Constraint Programming}: is a method specific
to \SMPT{} in the case where nets are ``fully reducible'' (the reduced
net has only one marking). In this case, reachable markings are
exactly the solution of the reduction equations ($E$) and verdicts are
computed by solving linear system of equations. \smallbreak
    
\noindent (8)\ \textbf{Enumeration}: performs an exhaustive
exploration of the state space and relies on the \textsc{Tina}
model checker. It can be used as a fail-safe, or to check the
reliability of our results. \smallbreak

%%%%%%%%%%%%%%%%%%%%%%%%%%%%%%%%%%%%%%%%%%%%%%%%%%%%%%%%%%%%%%%%%%%%%%%%%%%%%%%
%%%%%%%%%%%%%%%%%%%%%%%%%%%%%%%%%%%%%%%%%%%%%%%%%%%%%%%%%%%%%%%%%%%%%%%%%%%%%%%

\section{Commands, Basic Usage and Installation}
\label{sec:usage}

\SMPT{} requires Python version 3.7 or higher.
The easiest method for experimenting with the tool is to directly run
the \texttt{smpt} module as a script, using a command such as
\verb!python3 -m smpt!. %
Our repository includes a script to simplify the installation of the
tool and all its dependencies. It is also possible to find disk images
with a running installation in the MCC website and in artifacts
archived on Zenodo~\cite{zenodofm,zenodoamat}.
As usual, option \verb!--help! returns an abridged description of all
the available options. We list some of them below, grouped by
usage.\smallbreak

\noindent\textbf{Input formats.}
We accept Petri nets described using the Petri Net Markup Language
(PNML) \cite{hillah2010pnml} and can
also support colored Petri nets (using option \verb!--colored!) by
using and external unfolder \cite{mcc_colored}.
For methods that rely on polyhedral reductions, it is possible to
automatically compute the reduction (\verb!--auto-reduce!) or to
provide a pre-computed version (with option\linebreak
\verb!--reduced-net <path>!). It is also possible to save a copy of
the reduced net with the option \verb!--save-reduced-net <path>!. \smallbreak

\noindent\textbf{Verification methods.}
We support the verification of three predefined classes of safety
properties:
\emph{deadlock detection} (\verb!--deadlock!), which is
self-descriptive;
\emph{quasi-liveness} (\verb!--quasi-liveness t!), to check if it is
possible to fire transition \verb!t!; and \emph{reachability}
(\verb!--reachability p!), to check if there is a reachable marking
where place \verb!p! is marked (it has at least one token). It is also
possible to check the reachability of several places, at once, by
passing a comma-separated list of names,
\verb!--reachability p1,...,pn!; and similarly for liveness.
Finally, \SMPT{} supports properties expressed using the MCC property
language \cite{mcc_property_nodate}, an XML format encoding the class
of formulas described in Sect.~\ref{sec:theory}. Several properties
can be checked at once.
\smallbreak

\noindent\textbf{Output format.}
Results are printed in the text format required by
the MCC, which is of the form \texttt{FORMULA <id>
  (TRUE/FALSE)}. There are also options to output more information:
\verb!--debug! to print the SMT-LIB input/output code exchanged with the SMT
solver; \verb!--show-techniques!, to return the methods that successfully
computed a verdict; \verb!--show-time!, to print the execution time per
property; \verb!--show-reduction-ratio!, to get the reduction ratio;
\verb!--show-model!, to print the counterexample if it exists;
\verb!--check-proof!, to check verdict certificates (when we have one);
\verb!--export-proof!, to export verdict certificates (inductive invariants,
traces leading to counterexamples, etc.).  \smallbreak

\noindent\textbf{Tweaking options.}
We provide a set of options to control the behaviour of our
verification jobs scheduler. We can add a timeout, globally
(\verb!--global-timeout! \verb!<int>!) or per property
(\verb!--timeout <int>!). We can also restrict the choice of
verification methods (\verb!--methods <method_1> .... <method_n>!).
% which is useful for benchmarking.
Finally, option \verb!--mcc!  puts the tool in ``competition mode''.

%%%%%%%%%%%%%%%%%%%%%%%%%%%%%%%%%%%%%%%%%%%%%%%%%%%%%%%%%%%%%%%%%%%%%%%%%%%%%%%
%%%%%%%%%%%%%%%%%%%%%%%%%%%%%%%%%%%%%%%%%%%%%%%%%%%%%%%%%%%%%%%%%%%%%%%%%%%%%%%

\section{Comparison with Other Tools}
\label{sec:mcc}

We report on some results obtained by \SMPT{},
\textsc{ITS-Tools}~\cite{its_tools,thierry-mieg_structural_2020}, and
\textsc{Tapaal}~\cite{tapaal} during the 2022 edition of the
MCC~\cite{mcc:2022}.
We created a public repository~\cite{mccreach:2022} containing the
scripts used to generate the statistics and oracles for the 2022 edition of the Model
Checking Contest for the Reachability category.

\SMPT{} provides a default competition mode that implements a basic
strategy that should be effective in the conditions of the
MCC.
Basically,
we start by running the Random Walk and State Equation methods in parallel with a timeout of $\SI{120}{\second}$, on all
formulas, in order to catch easy counterexamples and invariants as quickly as
possible. Then we run more demanding methods: BMC,
$k$-induction, PDR,
etc.
The rationale is that queries used in the reachability competition are
randomly generated and usually exhibit a bias towards
``counterexamples'' (CEX), meaning false AG properties or true EF
ones. Also, when the formula is an invariant (INV), for instance a
``true AG property'', it can often be decided with the State Equation
method.\smallbreak

Our tool is quite mature. It achieved a perfect reliability score (all
answers are correct) and ranked at the third position, behind
\textsc{Tapaal} and \textsc{ITS-Tools}.
We display the results in a
Venn diagram where we make a distinction between CEX and INV
properties. There is a total of $50\,187$ answered queries (with
almost $60\%$ CEX). We observe that a vast majority of these queries
($41\,006$) are computed by all tools, and can be considered
``easy''. Conversely, we have $9\,181$ difficult queries, solved by
only one or two tools.

\begin{figure}[htb]
  \centering
  \makebox[\textwidth][c]{\includegraphics[width=\textwidth]{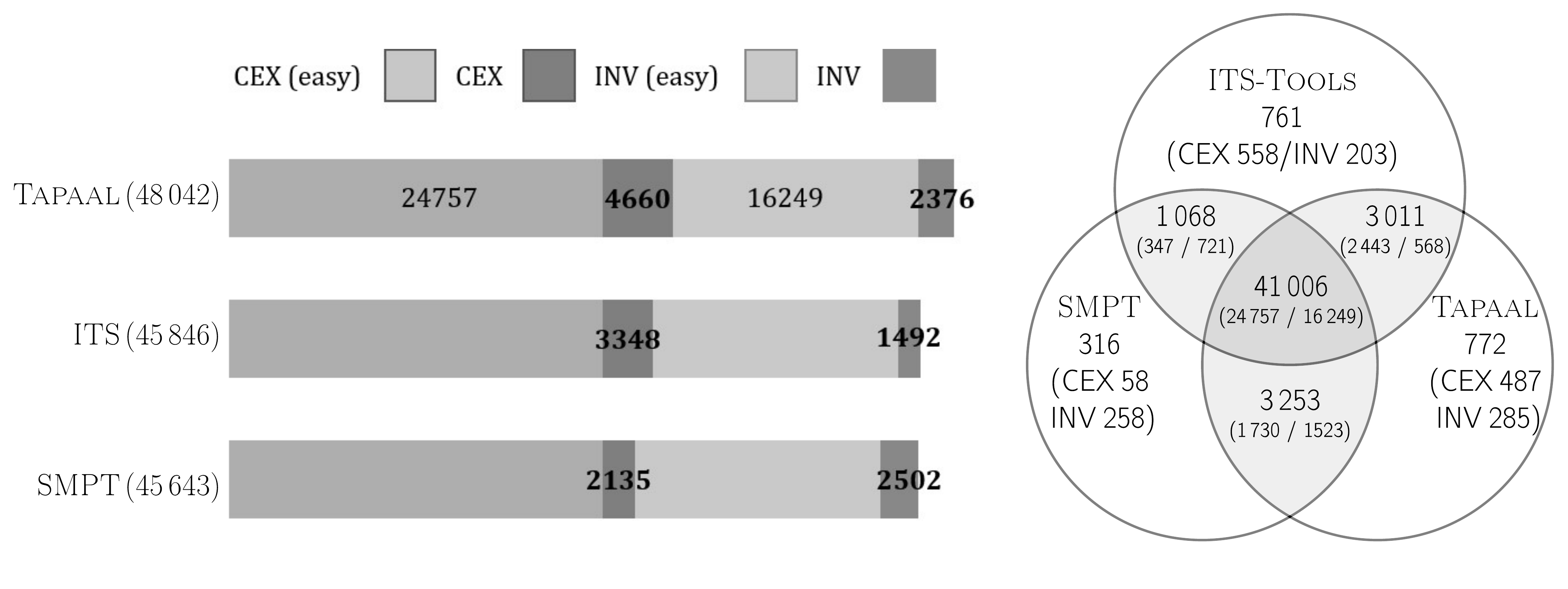}}
  \caption{Comparison of tools on all computed queries}
  \label{fig:venn}
\end{figure}

We also provide a bar chart where we distinguish between
easy/difficult, and CEX/INV queries. We observe that, while \SMPT{}
ranks last in the number of unique queries, it behaves quite well with
invariants (INV); which is the category we target with our most
sophisticated methods.
Overall, we observe that \SMPT{} performs well compared to other
state-of-the-art tools in the Model Checking Contest and that it is a
sensible choice when we try to check invariants.

%%%%%%%%%%%%%%%%%%%%%%%%%%%%%%%%%%%%%%%%%%%%%%%%%%%%%%%%%%%%%%%%%%%%%%
%%%%%%%%%%%%%%%%%%%%%%%%%%%%%%%%%%%%%%%%%%%%%%%%%%%%%%%%%%%%%%%%%%%%%%

\section{Future Work}

Work on \SMPT{} is still ongoing. At the moment, we concentrate on
methods to quickly discover counterexamples. The idea is to combine
polyhedral reductions and random exploration in order to find
counterexamples directly in the reduced net.
We also plan to improve our use of the ``state equation'' method, in
particular by identifying new classes of Petri nets for which all
potentially reachable markings are indeed reachable. A problem we
already started to study in a different
setting~\cite{hujsa2020checking}.

%%%%%%%%%%%%%%%%%%%%%%%%%%%%%%%%%%%%%%%%%%%%%%%%%%%%%%%%%%%%%%%%%%%%%%
%%%%%%%%%%%%%%%%%%%%%%%%%%%%%%%%%%%%%%%%%%%%%%%%%%%%%%%%%%%%%%%%%%%%%%

% \newpage

% %%%%%%%%%%%%%%%%%%%%%%%%%%%%%%%%%%%%%%%%%%%%%%%%%%%%%%%%%%%%%%%%%%%%%%%%%%%%%%%
% %%%%%%%%%%%%%%%%%%%%%%%%%%%%%%%%%%%%%%%%%%%%%%%%%%%%%%%%%%%%%%%%%%%%%%%%%%%%%%%

% \newpage
\bibliographystyle{splncs04}
\bibliography{bibfile}

\end{document}